\documentstyle[12pt]{article}
\begin{document}
\newcommand{\D}{{\cal D}}
\newcommand{\tr}{\bigtriangleup}
\newcommand{\rv}{\stackrel{\leftarrow}
{\frac{\delta }{\delta \eta}}}
\newcommand{\rvs}{ * \stackrel{\leftarrow}
{\frac{\delta }{\delta \eta}}}
\newcommand{\vv}[2]{\frac{\,\delta #1}{\delta  #2 }}
\newcommand{\vvv}[3]{\frac{\delta^2 #1}{\delta  #2
                     \,\delta #3 }}
\newcommand{\bin}[2]{\left(\!\!\begin{array}{c}#1\\#2
                     \end{array}\!\!\right)}
\newcommand{\free}{ m^2 - \partial\,^2}
\newcommand{\vgN}{\{ G^N\}}
\newcommand{\lbd}{\lambda}
\newcommand{\lb}{\label}
\newcommand{\be}{\begin{equation}}
\newcommand{\ee}{\end{equation}}
\newcommand{\ba}{\begin{eqnarray}}
\newcommand{\ea}{\end{eqnarray}}
\setlength{\unitlength}{1mm}
\thicklines

\title{{\Large\bf Singular Perturbations in Quantum Field Theory}\thanks{The
work is supported in part by the RFFI grant no. 95-02-03704a}}

\author{\large V.E.~Rochev and P.A.~Saponov}
\date{\it Institute for High Energy Physics, Protvino, Russia}
\maketitle

\begin{abstract}

In this talk we discuss a new approximation scheme for non-perturbative
calculations in a quantum field theory which is based on the fact
that the Schwinger equation of a quantum field model belongs
to the class of singularly perturbed equations.
The self-interacting scalar field and the Gross-Neveu model
are taken as the examples and some non-perturbative solutions
of an equation for  the propagator are found for these models.
The application to QCD is also discussed.

\end{abstract}

\section{Preliminaries}

It is well known  that in solving an equation approximately one
should distinguish two kinds of perturbations: singular and regular ones.

A perturbation is called to be singular if the term neglected at the
leading approximation is (in some sense) the main term for a  given
equation. For example, in the case of differential equation such a
term  contains the highest derivative.

The perturbative solution for the singularly perturbed system can have
nothing in common with the true solution of a given problem.
As an elementary example, let us consider the following differential
equation:
\be
\lbd\,{\dot x} = t - x \lb{1}
\ee
supplied with the boundary condition:
$x(0) = X$. The perturbative series for this problem consists of
two terms
$$
x_{pert}\,(t) \equiv \sum \lbd^n x_n = t - \lbd.
$$
Though $x_{pert}$ is an exact solution of differential equation
(\ref{1}) it does not give a solution of this problem if $X \not= -\lbd$.
Indeed if we interested in values of derivatives of $x(t)$ at the
origin ("vacuum expectation values") we can easily find from (\ref{1})
that
$$
{\dot x}(0)=-{1 \over \lbd}X,\quad {\ddot x}(0)={1 \over \lambda^2}X +
{1 \over \lambda}, \quad \mbox{etc.}
$$
That is the derivatives at the origin has definitely nonperturbative
character when $X\not= -\lbd$ and therefore cannot be described
by the perturbative series in $\lbd$.
Such a situation is a consequence of an essential singularity
in $\lbd$ which enters the solution of this problem
$$
x(t) = t - \lbd + (X + \lbd)\,\exp \left(- \frac{t}{\lbd}\right).
$$

Many physical problems belong to the singularly perturbed class.
One of the well known examples is the problem of the flowing
of a viscous liquid near a boundary.

This process is described by the Navier-Stokes equation:
$$
\nu\,{\vec\nabla^2}{\vec v} = ({\vec v}{\vec\nabla})\,{\vec v}
+ \frac{1}{\rho}\, {\vec\nabla}p
$$
with the boundary condition fixed on the surface of the body
being flowed around:
$$
{\vec v}|_{\mbox{\footnotesize surf}} = 0.
$$

Note that the viscosity $\nu$ is the coefficient at the highest
derivative $\vec\nabla^2$. When $\nu = 0$ this equation goes into the
Euler equation for the ideal liquid. At first sight it seems to be
quite natural to take the Euler equation as a leading approximation to the
case of small $\nu$. However, solutions of the Euler equation cannot satisfy
the boundary condition for the viscous liquid. The same is true for the
perturbative theory, based on these solutions. The behaviour of the ideal
liquid nearby the boundary is governed by weaker condition:
$$
\left.{\vec v}_{\bot}\right|_{\mbox{\footnotesize surf}} = 0.
$$
That is why one should take into account the highest derivative term
of Navier-Stokes equation from the very beginning, even if the viscosity
$\nu$ is small.

Any quantum field theory model with interaction is also a typical
example of singularly perturbed system. Let us turn to the simplest
model of a scalar field $\phi(x)$, $x\in E_d$ with quartic selfinteraction:
$$
{\cal L}_{int} =-\frac{\lbd}{4!}\,\phi^4.
$$

In the quantum field theory one calculates the vacuum expectation values
($n$-point functions)
$$
G_n(x_1\dots x_n) = \raisebox{0.2ex}{$<$}0\raisebox{0.2ex}{$|$}
T\,\phi(x_1)\dots\phi(x_n)\raisebox{0.2ex}{$|$}0\raisebox{0.2ex}{$>$}.
$$
These functions is the derivatives of the generating functional
$$
G(j) = \sum \frac{1}{n!} G_n \,j^n
$$
calculated at $j=0$:
$$
G_n = \left.\vv{^nG}{j^n}\right|_{j=0}.
$$

The generating functional can be found as a solution of the Schwinger
equation:
$$
\frac{\lbd}{3!}\, \vv{^3G}{j^3} + (\free)\,\vv{G}{j} -j\,G = 0.
$$
so as $\lbd$ is the coefficient at the highest derivative term,
the perturbation theory in $\lbd$ is singular. The same is true
for any QFT model with interaction.

In the present report we discuss the new approximation scheme for the
Schwinger equation which would take into account the singular character
of this equation.
The proposed scheme approximates the nonperturbative solutions of the Schwinger
equation already at few first steps without summing expansions. This opens new
possibilities for the description of essentially nonperturbative phenomena
such as spontaneous symmetry breaking and others.
For more detailed consideration of technical problems
the readers are referred to \cite{RochevSap} (see also \cite{RochevS2}).

\section{Approximaton scheme}

One of the key problems in finding a non-perturbative solution
for the Schwinger equation is the problem of additional boundary
conditions \cite{RochevS2,BeCoSi,Ro}. Indeed, the Schwinger equation
being the differential equation of the order higher than one requires
several boundary conditions in order to fix its solution uniquely.
But generally one knows only one of them: the normalization
of generating functional $G(0) = 1$. The other boundary conditions
will determine the frames of physical phenomena which can be described
by the corresponding solution of the Schwinger Equation. So as we do not
know the explicit form of the additional boundary conditions we have to
impose some constraints on the solutions of the Schwinger equation
which would play the role of boundary conditions. In our approach such
a constraint is the connected structures correspondence principle.

In the light of all mentioned above the perturbation theory acquires
a peculiar role in the quantum field theory. Due to the Schwinger
equation belongs to the class of singularly perturbed equations the
iteration procedure of the perturbation theory can be closed without
additional boundary condition. Fixing  the generating functional norm
is sufficient for the perturbation theory. As a result the perturbative
solution fails to describe the physical phenomena which require the
boundary conditions different from those automatically given
by the iterative procedure of the perturbation theory. Nevertheless
up to now the perturbation theory is the only universal tool for
calculations in the quantum field theory.

So we will construct our scheme basing on the following requirements:
\begin{itemize}
\item[i)] The perturbative expansion in $\lbd$ is always one of the
possible solutions of the scheme;
\item[ii)] Topological properties (connectivity) of the approximant
to be found correspond to those of perturbative one.
\end{itemize}
The last requirement guarantees the true connected structure of the
approximant and serves as an additional boundary condition which is
necessary for obtaining a close system of the Dyson equations for
the lowest Green functions.

Discussed below is the scheme for the theory $\lbd\,\phi^4_d$. The
generalization to any other QFT model is rather obvious and does not contain
principle difficulties.

Introduce the perturbative approximant
$$
G^N_{pert} = \sum_{n=0}^N G^{(n)}, \qquad G^{(n)} = O(\lbd^n),
$$
which can be calculated in the standard way and {\em define} the
approximant $G^N$ as a solution of the equation:
\be
\frac{\lbd}{3!}\,\vv{^3G^N}{j^3} + (\free)\,\vv{G^N}{j} - j\,G^N =
\frac{\lbd}{3!}\,\vv{^3G_{pert}^N}{j^3} + (\free)\,\vv{G_{pert}^N}{j}
- j\,G_{pert}^N.
\lb{2}
\ee
When $N\rightarrow\infty$, $G^N_{pert}\rightarrow G_{pert}\equiv
\sum_{n=0}^{\infty} G^{(n)}$.  $G_{pert}$ is an exact solution of
the Schwinger equation therefore due to (\ref{2}) $G^N$ also tends to
some exact solution of the Schwinger equation.

In accordance with the requirement  ii) we will use the connected
structures correspondence principle as the additional boundary condition
for this equation.
Namely, we will require the correspondence of the connected structures of
$G^N$ and $G^N_{pert}$.

To be more precise we should consider some set of approximants $\vgN$
which obey
equation (\ref{2}) and connected structures correspondence principle. It is
obvious that $G^N_{pert}\in\vgN$ at any rate, therefore our scheme will be
nontrivial if $\vgN/G^N_{pert}\not=\emptyset$.

The choice of the proper solution should be made on the base of additional
physical requirements such as the minimum of the ground state energy etc.

The connected structure of $n$-point function can be found
from the theorem on the connectivity of the logarithm. A functional
$$
 Z(j) = \log\, G(j)
$$
is a generating functional for the connected Green functions:
$$
 Z_n =\left. \vv{^nZ}{j^n}\right|_{j=0} = (G_n)^{con}.
$$
For example
$$
G_4 = 3\tr\tr + G_4^{con},
$$
where $G_4$ is the four-point function and $\tr$ is the {\em full}
propagator:

\begin{picture}(100,15)(10,8)

\put(25,15){\circle{10}} \put(23,13.5){$G_4$}
\put(14.5,19){\line(1,0){7.5}}
\put(14.5,11){\line(1,0){7.5}}
\put(28,19){\line(1,0){7.5}}
\put(28,11){\line(1,0){7.5}}

\put(38,14.5){=}

\put(44,19){\line(1,0){12}}
\put(44,11){\line(1,0){12}}

\put(58.5,14.5){+}

\put(63,19){\line(1,0){4}}
\put(67,19){\line(1,-1){8}}
\put(75,11){\line(1,0){4}}
\put(63,11){\line(1,0){4}}
\put(67,11){\line(1,1){8}}
\put(75,19){\line(1,0){4}}

\put(81,14.5){+}

\put(85,15){\oval(14,8)[r]}
\put(101,15){\oval(14,8)[l]}

\put(103,14.5){+}

\put(118.5,15){\circle{10}} \put(114.5,13.5){$G_4^{con}$}
\put(108,19){\line(1,0){7.5}}
\put(108,11){\line(1,0){7.5}}
\put(121.5,19){\line(1,0){7.5}}
\put(121.5,11){\line(1,0){7.5}}

\end{picture}

For the six-point function we have:
$$
G_6 = 15 \tr\,\tr\,\tr + 15\tr\,G_4^{con} + G_6^{con}
$$

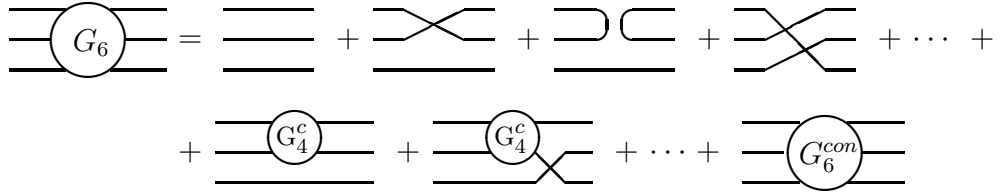
\begin{picture}(100,28)(10,19)

\put(25,40){\circle{10}} \put(23,38.5){$G_6$}
\put(14.5,44){\line(1,0){7.5}}
\put(14.5,36){\line(1,0){7.5}}
\put(28,44){\line(1,0){7.5}}
\put(28,36){\line(1,0){7.5}}
\put(14.5,40){\line(1,0){5.5}}
\put(30,40){\line(1,0){5.5}}

\put(37,39){=}

\put(43,44){\line(1,0){12}}
\put(43,40){\line(1,0){12}}
\put(43,36){\line(1,0){12}}

\put(58,39){+}

\put(63,44){\line(1,0){4}}
\put(67,44){\line(2,-1){8}}
\put(75,40){\line(1,0){4}}
\put(63,40){\line(1,0){4}}
\put(67,40){\line(2,1){8}}
\put(75,44){\line(1,0){4}}
\put(63,36){\line(1,0){16}}

\put(82,39){+}

\put(87,42){\oval(14,4)[r]}
\put(103,42){\oval(14,4)[l]}
\put(87,36){\line(1,0){16}}

\put(106,39){+}

\put(111,44){\line(1,0){4}}
\put(111,40){\line(1,0){4}}
\put(111,36){\line(1,0){4}}
\put(115,40){\line(2,1){8}}
\put(115,36){\line(2,1){8}}
\put(115,44){\line(1,-1){8}}
\put(123,44){\line(1,0){4}}
\put(123,40){\line(1,0){4}}
\put(123,36){\line(1,0){4}}

\put(130,39){+\ $\cdots$ \ +}

\put(37,24){+}

\put(52.5,27){\circle{7}} \put(50,26){{\footnotesize G}$_4^{c}$}
\put(42,29){\line(1,0){7.5}}
\put(42,25){\line(1,0){7.5}}
\put(55.5,29){\line(1,0){7.5}}
\put(55.5,25){\line(1,0){7.5}}
\put(42,21){\line(1,0){21}}

\put(66,24){+}

\put(81.5,27){\circle{7}} \put(79,26){{\footnotesize G}$_4^{c}$}
\put(71,29){\line(1,0){7.5}}
\put(71,25){\line(1,0){7.5}}
\put(84.5,25){\line(1,-1){4}}
\put(88.5,21){\line(1,0){3.5}}
\put(84.5,29){\line(1,0){7.5}}
\put(71,21){\line(1,0){13.5}}
\put(84.5,21){\line(1,1){4}}
\put(88.5,25){\line(1,0){3.5}}

\put(95,24){+\ $\cdots$\ +}

\put(123,25){\circle{10}} \put(119.5,23.5){$G^{con}_6$}
\put(112,29){\line(1,0){7.5}}
\put(112,21){\line(1,0){7.5}}
\put(126,29){\line(1,0){7.5}}
\put(126,21){\line(1,0){7.5}}
\put(112,25){\line(1,0){5.5}}
\put(128,25){\line(1,0){5.5}}

\end{picture}

\noindent
These relations are of the general type for the full Green functions.

The connected structure of the perturbative approximant possesses the
following property at  $O(\lbd^N)$ order of approximation:
\begin{eqnarray*}
\left( G_{2n}^N\right)^{con}_{pert} \not= 0 &\mbox{if}& n\leq N+1,\\
\left( G_{2n}^N\right)^{con}_{pert} = 0 &\mbox{if}& n> N+1.\\
\end{eqnarray*}
(In the model involved $G_{2n+1} = 0$ due to the parity conservation.)
For example, at the order $O(1)$ (the free theory)
\begin{eqnarray*}
\left( G_{2}^0\right)^{con}_{pert} & = &\tr_c = (\free)^{-1},\\
\left( G_{2n}^0\right)^{con}_{pert}& = & 0\quad  n>1.\\
\end{eqnarray*}
At the order $O(\lbd)$ we have
$$
(G_2^1)^{con}_{pert}\not=0,\quad (G_4^1)^{con}_{pert}\not=0\quad\mbox{and}
\quad (G^1_{2n})_{pert}^{con} = 0\mbox{\ if\ }\ n>2, \mbox{\ etc.}
$$
This property will be the base of the connected structures correspondence
principle. Namely let us require at the $N$-th step of approximation that
$$
\left( G_{2(N+2)}^N\right)^{con} = 0.
$$
This equation allows us to obtain the relation:
\be
G^N_{2(N+2)} = F\left[\tr,\,G_4^N,\,\dots,G^N_{2(N+1)}\right].\lb{3}
\ee
With the help of (\ref{3}) we can close the system of $N+1$
Dyson equations for $N+1$ lowest functions $\tr,\ G_4^N,\ \dots
G^N_{2(N+1)}$ at the $N$-th step of approximation.
For example, at the first step (N=0)
$$
\left(G_4^0\right)^{con} = 0 \Rightarrow G_4^0 = 3\tr^0\tr^0
$$
and we have one equation for the propagator $\tr^0$.
At the second step (N=1)
$$
\left(G_6^1\right)^{con} = 0 \Rightarrow G_6^1 = 15 \tr^1\tr^1\tr^1
+ 15 \tr^1\left(G_4^1\right)^{con}
$$
and we have the system of two equations for the propagator $\tr^1$
and the two-particle function $G_4^1$ and so on.

At any step of the scheme the system of these equations has at least
one solution (the perturbative approximant). All other solutions
(if they exist) are nonperturbative ones.

\section{The first step: an equation for the propagator.}
In this section we reproduce only the list of results for some
models. Detailed derivation can be found in \cite{RochevSap}.

{\it 1. The theory $\phi^4_d$ in Euclidean metric.} \par

An equation for the propagator $\tr^0\equiv \tr$ at the leading
approximation has the form:
\be
(\free)\,\tr + \frac{\lbd}{2}\, \tr\,\tr = 1 +\frac{\lbd}{2}\, \tr_c\,\tr_c.
\lb{4}
\ee
Besides the trivial perturbative solution $\tr = \tr_c$ there exist
nonperturbative ones:

i) \underline{$ d=2$.}\par
\be
\tr = \frac{1}{p^2 + \mu^2} + O(\lbd),\lb{5}
\ee
where $\mu^2 \simeq m^2\exp\left(-\frac{8\pi m^2}{|\lbd|}\right),
\lbd\rightarrow -0$.

ii) \underline{$d = 4$.}\par
There exist the solution of the form
\be
\tr = \frac{1}{\lbd}\, \frac{1}{p^2 + \mu^2} + O(1), \quad \mu^2 =
O\left(\frac{1}{\lbd}\right),\quad \lbd \rightarrow +0,\lb{6}
\ee
and the dipole solution
\be
\tr = \frac{A}{p^2 + m^2} + \frac{B}{(p^2 + m^2)^2} \qquad B = o_{\lbd}(A).
\ee
\vskip 5 mm
{\it 2. The Gross-Neveu model.}\par
This is a model of a spinor field  $\psi(x)$ in the two dimensional
Minkowsky space with the Lagrangian:
$$
{\cal L} = \bar\psi\,(i\hat\partial - m)\,\psi + \frac{\lbd}{2}\,
(\bar\psi\psi)^2.
$$
The equation for the propagator $S^0 = S$ reads as follows:
\be
(m - i\hat\partial)\,S + i\lbd\,S\,S - i\lbd\,(TrS)\,S =
\mbox{\bf 1} + i\lbd\, S_c\,S_c - i\lbd\,(TrS_c)\,S_c.
\ee
Here $ S_c = (m - i\hat\partial)^{-1}$ is the free propagator.

Besides the trivial solution $S = S_c$ there exist nonperturbative
solutions. At chiral limit $m\rightarrow 0$ they  correspond to the
spontaneous breakdown of the chiral symmetry. We would like to point
out that in the original paper by Gross and Neveu the limit $N_c \rightarrow
\infty$ for the $N_c$-component field was considered. In our approach the
spontaneously broken solution exists for the one component field at small
$\lbd$. This solution has the form
\be
S(p) = \frac{4\pi}{\lbd}\,\frac{{\hat p}  + \mu}{\mu^2 - p^2},\qquad
\mu = O\left(\frac{1}{\lbd}\right).
\ee
\vskip 5 mm

{\it 3. QCD}\par
The equations for the quark propagator and ghost propagator have
only trivial perturbative solutions at the first step of
approximation.

The equation for the gluon propagator takes the form:
\be
(D^c)^{-1}_{\mu\nu}\D_{\mu\nu} + i g^2 C\,(\D_{\mu\lbd}\D_{\lbd\nu}
- \D_{\lbd\lbd}\D_{\mu\nu}) = g_{\mu\nu} + i g^2 C\,(D^c_{\mu\lbd}
D^c_{\lbd\nu} - D^c_{\lbd\lbd}D^c_{\mu\nu}),\lb{10}
\ee
where $D^c_{\mu\nu}$ is the free propagator and ${\cal D}_{\mu\nu}$
is the one to be sought for. The constant $C$ is defined to be
$\delta^{ab} C = f^{anc}f^{bnc}$.
Besides the trivial solution $\D = D^c$ this equation has
nonperturbative ones and among them the well-known "dual superconducting"
solution of $1/p^4$ form.
However, all these solutions are forbidden by the gauge invariance
requirement. This is quite natural because at the first step only
the four-gluon vertex contributes into the equation (\ref{10}).
One can expect, that nontrivial and physically interesting solutions
(including the one with spontaneously broken chiral symmetry) will
appear at the next step of approximation when three gluon, quark-gluon
and ghost vertices will also play the game.

\section{Conclusion}
In conclusion we would like to emphasize once more that our scheme is
not a variant of the perturbation theory partial summation. The perturbation
theory (even being somehow summarized) cannot in principle exhaust
the full information contained in the equations of the quantum field theory.
The similar conclusions were made in the work \cite{Garsia} where the problem
of nonunique solution to the Schwinger equation was approached from the
different positions.
So we hope that our scheme (or similar to it) can be useful in describing
the nonperturbative content of quantum field models.

In this connection it is worth mentioning some ways for further development.
First of all it would be interesting to find a nontrivial solution
for the second step of the scheme which include a system of equations
for the propagator and connected part of the full vertex
(see \cite{RochevSap}). Another crucial point of this approach is to
develop criteria which could help to choose the proper solution from
all the  possible ones given by the scheme equations.
And, of course, the main interest is to try to describe the nonperturbative
phenomena in realistic physical theories such as QCD.

\end{document}